# Realistic UE Antennas for 6G in the 3GPP Channel Model

Simon Svendsen, Dimitri Gold, *Senior Member, IEEE*, Christian Rom*, Volker Pauli,* and Vuokko Nurmela[1]

## ABSTRACT

The transition to 6G has driven significant updates to the 3GPP channel model, particularly in modeling UE antennas and user-induced blockage for handheld devices. The 3GPP Rel.19 revision of TR 38.901 introduces a more realistic framework that captures directive antenna patterns, practical antenna placements, polarization effects, and element-specific blockage. These updates are based on high-fidelity simulations and measurements of a reference smartphone across multiple frequency ranges. By aligning link- and system-level simulations with real-world device behavior, the new model enables more accurate evaluation of 6G technologies and supports consistent performance assessment across industry and research.

## INTRODUCTION

The transition from 5G to 6G presents a rare opportunity to validate and improve how we model the channel and specifically user equipment (UE) in wireless system simulations. In 3GPP, 5G Rel.19 has been paving the way towards 6G. Rel.19 study item (SI) on channel modelling enhancements [1] targeted the validation of the existing models in Rel.18 version of technical report (TR) 38.901 [2] using new measurements in 7–24 GHz band. A general overview of the updated 3GPP channel model is provided in [3].

In the new Rel.19 version of the TR 38.901 [4], the standard UE antenna model was fundamentally revised to address shortcomings of the legacy model. The previous 5G UE antenna model (defined back in Rel.15) was carried over from early LTE (Rel.8, circa 2008) and, while functional, it failed to capture many complexities of modern smartphones.

### LIMITATIONS OF THE LEGACY UE MODEL

The 5G UE antenna model made several simplifying assumptions that diverge from reality:

---

[1] All the authors are with Nokia. Simon Svendsen and Christian Rom are with Strategy & Technology, Nokia Standards, Aalborg 9220, Denmark. Dimitri Gold is with Strategy & Technology, Nokia Standards, Espoo 02610, Finland. Volker Pauli is with Strategy & Technology, Nokia Standards, Munich 81541, Germany. Vuokko Nurmela is with Nokia Mobile Networks, Espoo 02610, Finland.

**Radiation Pattern:** It used an isotropic pattern for the UE. Such radiator emits equally in all directions (0 dBi gain), removing any single antenna-specific effects in simulations. This can be valid for some theoretical studies but is not realistic for modern smartphones. It is an acceptable proxy for early-generation handsets with external antennas, but it's a poor fit for today's phones with internal antennas integrated into the device chassis. In short, real handset antennas are far from omnidirectional – they have directive lobes and nulls influenced by the phone's properties, which the Rel.15 model did not capture.

**Antenna Placement:** It assumed a uniform array-like placement of multiple uniform antennas, analogous to base station antenna arrays with half-wavelength element spacing. For example, 4×4 MIMO in the UE was modeled as a dual-polarized 1×2 array (4 elements) with ideal half-wavelength spacing [5]. While such an array model is reasonable for a fixed base station or router, it is not applicable to handheld devices. In real smartphones, antennas are irregularly distributed along the edges of the device with varying spacing and orientation. The coupling and radiation characteristics differ for each antenna depending on its location. The old model's one-size-fits-all array approach could not reflect these differences.

**Polarization:** It ignored realistic polarization behavior, simply assuming perfectly orthogonal polarizations for multiple antenna ports. The Rel.15 model treated the two ports of a MIMO pair as having ideal cross-polarized patterns with no coupling, regardless of direction. In a real phone, antennas do not have such idealized polarization isolation – each antenna often radiates a mix of polarizations, and no two elements maintain pure orthogonality over all angles. By neglecting this, the old model missed polarization mismatch and coupling effects present in actual devices.

**User Blockage:** TR 38.901 [2] included a very simplified "self-blockage" model to represent the user's hand or head blocking the signal. It defined a fixed attenuation region (e.g. a cone of certain angles) where the UE signal is weakened by a fixed amount (30 dB) for all antennas equally. While this accounted for blockage in a coarse way, it did not differentiate which antenna was covered by the hand – in reality, a user's grip affects some antennas much more than others [6]. The uniform blockage region led to inconsistent and unrealistic impacts on different antennas.

Given these limitations, it became clear that the 5G UE antenna model was inadequate for today's smartphones and emerging 6G use cases. The motivation behind the new Rel.19 model was to make link-level and system-level simulations (LLS and SLS) more representative of real-world device behavior. By aligning simulations closer to reality, the industry can gain more accurate insights into device and network performance, leading to better designs and optimizations in 6G.

## Overview of the New Model

3GPP's Rel.19 update to TR 38.901 [3] introduced a comprehensive new UE antenna model with several key enhancements to address the above gaps:

Realistic Form Factor and Antenna Locations: A standard handheld form factor of 150 × 70 mm is assumed, with a realistic multi-antenna layout around the device's perimeter. Up to eight antenna element positions are defined on this reference handset, capturing the typical locations of antennas at the top, bottom, and corners of a smartphone.

Directional Element Patterns: Instead of an isotropic point source, a directive 3D radiation pattern is specified for each UE antenna element. The reference pattern has about 5.3 dBi peak gain, a ~125° half-power beamwidth, and a front-to-back ratio of 22.5 dB. This pattern is derived to approximate a modern smartphone antenna's free-space radiation, and it is applied with appropriate orientation for each of the eight antenna positions (each pointing in a different direction). The model thus recognizes that smartphone antennas have lobes (not uniform radiation) and that each antenna "sees" the environment differently.

Polarization and Antenna Orientation: The new model explicitly includes two orthogonal polarization components (referred to as $\theta$ and $\phi$ in spherical coordinates) rather than assuming idealized polarization separation. The standard provides a method to rotate the reference pattern to any antenna's location and orientation on the device, computing the corresponding polarized components for that orientation. This ensures that the polarization of each element's radiation is realistically represented, including any tilt or mixing that occurs when antennas are mounted on a phone.

Antenna Blockage and Performance Variations: A new spatially varying blockage model was introduced to account for the user's hand/head effects more realistically. Instead of one blanket attenuation region for all antennas, the model applies different attenuation values to each antenna element when the device is in use, based on that antenna's likely exposure to blockage. These attenuation factors were derived from detailed simulations for up to 8.4 GHz carrier and measurements around 15 GHz with anthropomorphic hand and head phantoms, so they correlate with actual observed losses when a user grips a phone.

The new model also captures such per-port gain imbalances – the fact that some antennas will perform better than others depending on their location and implementation.

By incorporating these features, the Rel.19 UE antenna model aligns much more closely with realistic smartphone behavior. In the following sections, we delve into each aspect of the new model – explaining how these changes were derived and why they matter for simulation accuracy and future device design.

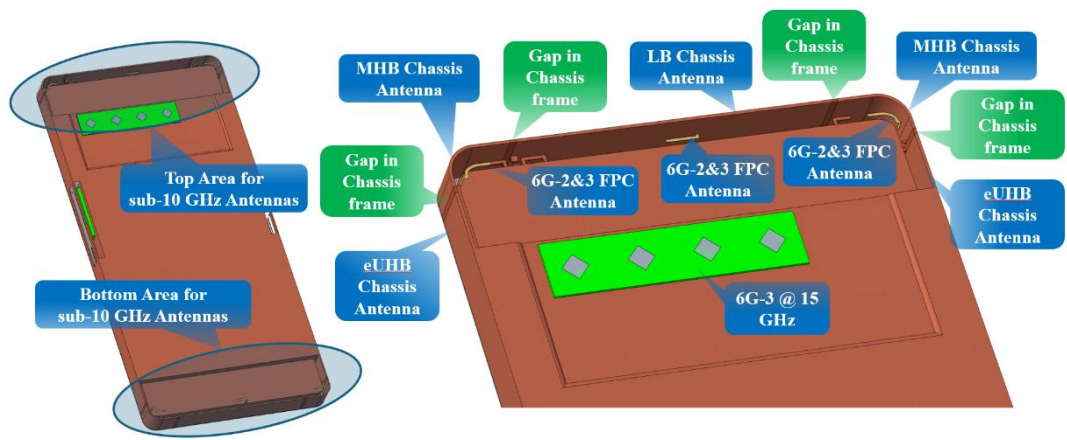

**FIGURE 1.** Illustration of the physical locations of the antennas of the realistic reference 6G smartphone model.

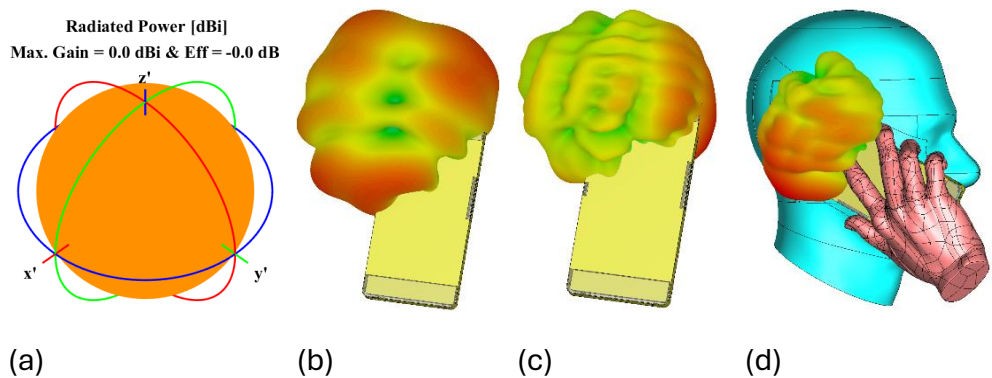

**FIGURE 2.** Single antenna radiation patterns: for ideal omnidirectional antenna (a), for reference phone at 3.8 GHz (b), and 7.4 GHz (c) in open space and with the presence of head and hand grip (d).

# UE Radiation Models: From Theory to Realism

Designing a better UE antenna model first requires understanding how real devices radiate, especially under the influence of users. To that end, advanced simulation tools and measurement data were used to generate reference data for the model development. A key contribution came from comprehensive electromagnetic (EM) simulations of a reference smartphone, which served as a proxy for a "typical" 6G-capable handset [7].

## Reference Smartphone Simulations

In the model development process, a detailed mechanical model of a smartphone was created. Considered 6G reference smartphone design had dimensions of ~150 × 70 mm, with 19 antenna elements integrated to cover a wide range of frequencies from 0.6 GHz up to 15.4 GHz. These included multiple antennas for low bands (LB: 600–960 MHz), mid-high bands (MHB: 1.7–2.7 GHz), extended ultra-high bands (eUHB: including the new 6G bands 1: 4.4-4.8 GHz, 2: 7.1-8.4 GHz, and 3: 14.8-15.4 GHz) positioned at various locations of the device, as shown in Fig. 1. Using this reference device model, full 3D EM simulations were performed to calculate radiation patterns and efficiencies for each antenna under different conditions. State-of-the-art simulation software (such as CST Microwave Studio [8]) was employed, and the device was simulated in free space as well as with user phantoms. Fig. 2 illustrates how significantly omni-directional radiation pattern assumed in the past is different from realistic one.

Several use-case scenarios were defined to mimic how a phone is typically used:

**Free Space**: Phone by itself, no user influence.

**One-Hand Browsing**: A right-hand grip holding the phone in a browsing orientation as defined by Cellular Telecommunications and Internet Association (CTIA) standard test guidelines [9].

**Two-Hand Browsing**: A custom dual-hand grip where the user holds the phone with both hands (e.g., for texting or gaming in landscape mode).

**Head-and-Hand (Talk Mode)**: A CTIA scenario with a head phantom and a hand, simulating a phone held up to the right side of the head during a typical voice call.

Each of these scenarios introduces different blockage and detuning effects on the antennas. For example, a one-hand grip tends to cover the antennas in the lower part of the phone, whereas a head-and-hand scenario also adds attenuation to the antennas in the top of the phone (see Fig. 2-d).

By simulating all these cases, a range of realistic effects was captured. Moreover, these simulations were run at multiple frequencies to span across bands from 0.7 to 7.8 GHz. This ensured that frequency-dependent behavior (like different implementation loss and - antenna coupling at different frequencies) can be observed. To make simulations as realistic as possible, lossy materials and components were included in the phone model. These measures ensured that the simulated antenna efficiencies (total radiated power vs. input power) matched what is typically seen in real high-end smartphones.

## Key Findings from Simulations

The detailed simulations confirmed just how non-uniform real UE antenna behavior is, validating the need for a better model. For example, even in free space, a smartphone's multiple antennas do not provide equal coverage in all directions. Each antenna has a distinct radiation pattern with its own peaks and nulls, dictated by its location on the device and carrier frequency as shown in Fig.2-b&c. Simulations of four representative antennas on the reference phone showed gain imbalances often exceeding 10 dB between the best and worst antenna for a given direction. In other words, the phone exhibits a highly directional composite pattern – a far cry from the omni-directional assumption of the legacy UE model.

Crucially, when the user's hand comes into play, these disparities become even more pronounced. It was observed that the coverage imbalance can exceed 30 dB in certain orientations with a hand grip, meaning one antenna's signal may be practically obliterated by blockage while another antenna (uncovered) still has decent gain.

Another finding was that combining multiple antennas on a handset doesn't behave like a textbook antenna array. In theory, if there are two ideal uniform antennas in the UE, one might expect a 3 dB gain improvement when using them together (since the antenna aperture is doubled and twice the power is radiated or twice the energy is captured for a given angular direction). However, due to the non-uniform patterns and phases of real handset antennas, simply adding more antennas doesn't always yield the full 3 dB per doubling benefit. Simulations showed that the largest gain improvements often came from combining just a few antennas while some combinations will reduce the maximum gain value compared to a single antenna. This is because ideal power combining is not always possible for single feed non-uniform antennas where the gain is not in a single polarization as seen for dual polarized uniform antenna arrays. The new model takes this into account by treating each antenna element separately with its own polarization pattern, allowing realistic beamforming and antenna selection simulations that reflect these non-idealities.

In summary, the use of high-fidelity simulations and measurements was critical in shaping the Rel.19 UE antenna model. These tools provided quantitative backing for the new approaches – showing exactly how a phone's antenna performance varies with placement, frequency, polarization, and user grip. Armed with this data, 3GPP introduced new standardized assumptions in [4] to bring simulation models closer to reality.

# Reference UE Dimensions and Antenna Candidate Locations

One of the most visible changes in Rel.19 is the specification of multiple UE antenna positions and UE reference radiation pattern in the channel model. Instead of treating the UE as a single isotropic point, Section 7.3.0 of TR 38.901 [4] now defines a set of eight antenna element patterns positioned around a reference smartphone outline.

The reference device possesses a handheld form factor of 15 cm × 7 cm x 0 cm (for length x width x height) corresponding to the size of a typical smartphone. On this device, eight antenna element locations are designated, roughly corresponding to realistic placements on a phone's perimeter as illustrated in Fig. 3. This covers the common positions where manufacturers often place antennas to support multiple MIMO streams and various bands. In addition to the handheld UEs, the customer premises equipment (CPE), e.g., fixed wireless access devices, with reference form factor of 0 cm x 20 cm x 20 cm were considered in the study.

The radiation pattern at each candidate antenna location is assumed to face outward from the device, with its peak gain pointing in a distinct direction in the phone's local coordinate system. This arrangement ensures that all used antennas cover different parts of the sphere, mimicking how real phone antennas collectively provide near-omnidirectional coverage when considered together.

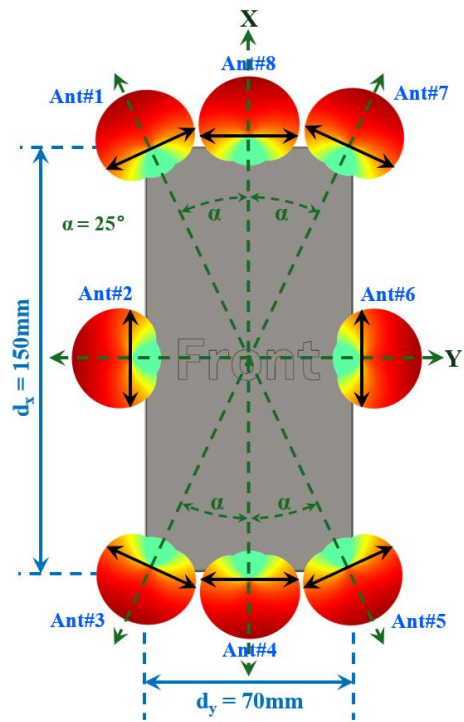

**FIGURE 3.** Rel.19 reference handheld UE with antenna candidate locations and orientations, top-down view.

Notably, the spacing between the UE antennas is non-uniform – some might be separated by the full device length (~150 mm), like a top vs. bottom antenna, while others (like two top-corner antennas) might only be ~70 mm apart. This irregular spacing and orientation means that if multiple antennas are used together, the combined pattern will exhibit frequency-dependent effects (because 70 mm is a different electrical spacing at 3 GHz versus 6 GHz, for instance).

## Radiation Pattern

The foundation in the new UE antenna model is a new reference radiation pattern $A''_{\text{dB}}(\theta'', \phi'')$. This pattern, defined in Tab. 7.3-2 of [4], represents the directive gain shape of a single smartphone antenna in free space. It is somewhat analogous to how base station antenna patterns were defined in [2], but tuned for handheld devices. The key characteristics of the reference pattern are:

- **A maximum directional gain of 5.3 dBi**. This is the peak gain of the main lobe, which is several dB higher than an isotropic radiator (0 dBi) but not as high-gain as a large antenna or array. It reflects the fact that a smartphone antenna does concentrate power in a particular direction.

- **A 3 dB beamwidth of approximately 125°** in both the elevation and azimuth cuts. This indicates the main lobe is fairly broad – spanning about a third of the sphere. The antenna does not narrowly beamform; it radiates over a wide angle, which is consistent with radiation patterns seen on smartphones.

- **A front-to-back ratio on the order of 22.5 dB.** As the main direction of the pattern has a gain of 5.3 dBi, other directions will need to have much lower gain to maintain the same antenna efficiency. This is modelled by attenuating radiation power in the direction opposite to the maximum gain by 22.5 dB.

These parameters were chosen based on compromise between different company proposals [10] to achieve realistic directivity observed in smartphone antennas around mid-band frequencies. Moreover, such parameter selection ensures 0 dB antenna efficiency. Therefore, the simulation results with new and legacy omnidirectional antenna patterns can be compared directly.

For each of the eight candidate antenna locations demonstrated in Fig. 3, the reference pattern should be rotated following the max gain direction marked with the dashed arrow oriented from the center of the device. By doing this, each antenna's pattern is realistically oriented, and collectively the set of selected antennas can provide near-360° coverage. This is a significant improvement: instead of one notional UE pattern, we now have multiple, and they are based on real antenna placement strategy as seen in today's phones.

This approach still balances realism with simplicity. All antennas share the same reference pattern shape. In reality, different antennas on a phone can have different intrinsic patterns. As can be seen from Fig. 4, the 3GPP radiation pattern is reasonably close to the antenna radiation patterns of reference smartphone described in the previous section. The curves in Fig.4 show a 2D-cut ($A''_{dB}(\theta'' = 90°, \phi'')$) of the simulated radiation patterns (normalized to a 0 dB efficiency) at different frequencies (solid lines) and the new 3GPP reference UE antenna pattern (dashed line).

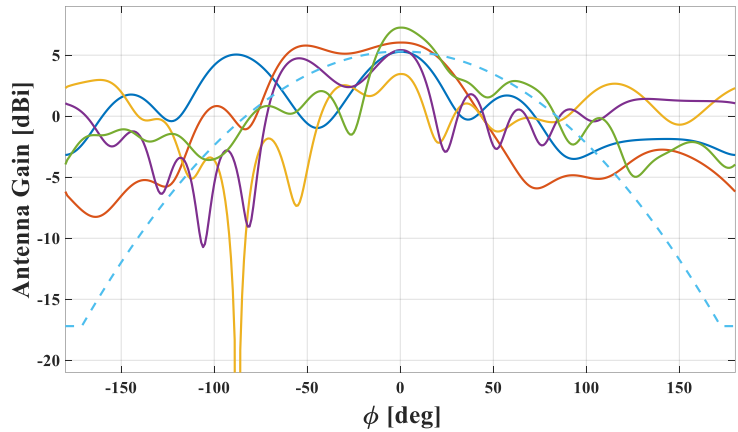

**FIGURE 4.** UE antenna radiation patterns for 3GPP model (dashed) and reference phone at different frequencies (solid).

Moreover, antenna imbalance is also allowed by the model. Randomized loss can be applied per UE antenna port. During 3GPP discussions [10], the proposed loss values were in the range from 2 to 3 dB, but no imbalance was agreed to be modelled by default.

For simulation users, the implication of the new model is that when setting up system-level studies (like evaluating MIMO or beamforming), the UE is no longer a black box isotropic node. Instead, one would instantiate a UE with multiple antenna ports, each with a specified gain pattern in 3D space. As the UE moves or rotates, the gain of each port in the direction of a serving cell will differ, and algorithms like beam selection or MIMO combining can be tested under those conditions.

In summary, Rel.19's introduction of multiple UE antenna locations and directive patterns marks a major step toward realism. It acknowledges that a smartphone isn't a point source – it's a collection of antennas with coverage that together approximate omnidirectional behavior, yet individually have distinct strengths and weaknesses. By standardizing these placements and patterns, 3GPP provides a common framework for industry and academia to evaluate new technologies (like 6G radios or advanced MIMO schemes) with assumptions that mirror a real handset's behavior.

## Polarization Components

Another critical aspect of the new UE antenna model is the treatment of polarization and the orientation of each antenna element's radiation pattern. Previous 3GPP models effectively ignored the intricacies of polarization on the UE side, assuming idealized dual-polarized channels with perfect isolation. In reality, smartphone antennas are typically single-feed elements that do not produce a purely vertically or horizontally polarized wave. Instead, the polarization of the radiated field varies with direction and often appears as a mix, i.e., as elliptical polarization shown in Fig. 5-a for one of the antennas on the reference smartphone. Blue ellipses correspond to left-handed and red - to right-handed polarizations. The Rel.19 model introduces a more nuanced approach to represent this, ensuring that simulations account for polarization mismatch and rotation effects.

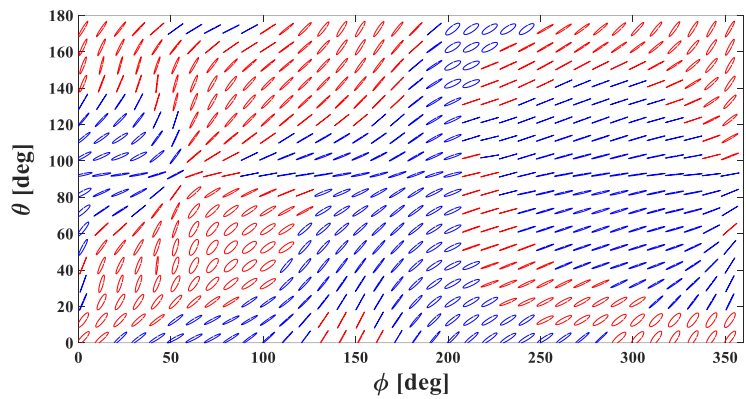

(a)

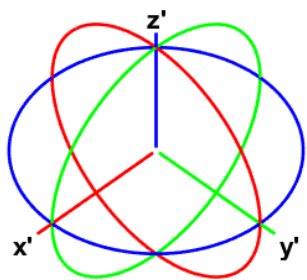 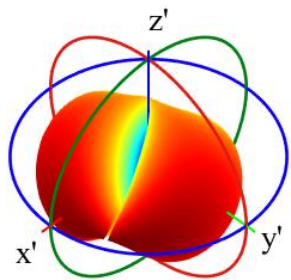

(b)          (c)

**FIGURE 5.** Illustration of polarization pattern for UE antenna #3 (a), and $\theta$ (above) and $\phi$ (below) polarization components for reference radiation pattern (b) and antenna #3 in LCS (c).

The standard defines a local UE local coordinate system (LCS) centered on the device as shown in Fig. 7.3-3 of [4] and in Fig. 3. This system can be referred to as a single-prime one. Each antenna element of the device is assigned an orientation in this UE LCS – basically, the $\alpha, \beta, \gamma$ angles that describe how its pattern is rotated in 3D relative to the orientation of reference pattern. However, specifying only the maximum gain direction is insufficient to fully define the antenna radiation pattern orientation and polarization should be also considered.

Instead of giving a single "total" gain pattern for each antenna, the model defines patterns for two orthogonal polarization components in spherical coordinate system. In simpler terms, one can think of them analogous to vertical and horizontal polarizations, but defined locally for the antenna's orientation. Initially, a reference radiation pattern is defined with all its gain in one polarization component $\theta$: $F''_{\theta''}(\theta'', \phi'') = \sqrt{A''_{dB}(\theta'', \phi'')}$ and $F''_{\phi''}(\theta'', \phi'') = 0$ for a reference double-prime antenna coordinate system (ACS). The orientation of the reference antenna radiation pattern within the UE LCS is shown in Fig. 5-b, where the maximum gain direction of the UE reference radiation pattern (x'') is aligned with the z' axis of the LCS.

Next, this pattern is rotated to a particular antenna's position/orientation in UE LCS. The polarization direction per antenna is aligned with the solid black arrows shown in Fig. 3. The standard uses a rotation matrix from equation (7.1-11) in [2] to compute how much of that pattern appears in the new $F'_{\theta'}(\theta', \phi')$ and $F'_{\theta'}(\theta', \phi')$ components for that orientation. Fig. 5-b illustrates how the reference pattern's power, initially all in $\theta''$, gets split into $\theta'$ and $\varphi'$ in LCS, for antenna location #3 after rotation.

Finally, the UE orientation in the global coordinate system (GCS) is defined by three 3D rotation angles $\Omega_{UT,\alpha}$, $\Omega_{UT,\beta}$, and $\Omega_{UT,\gamma}$ (see the examples in Fig. 7.3-4 and 7.3-5 from [4]). Correspondingly, the antenna polarization field patterns can be transformed from LCS to GCS using the same transformation following equation (7.1-11) in [2]. The derived filed components per each UE antenna $u$ termed $F_{rx,u,\theta}(\theta, \varphi)$ and $F_{rx,u,\varphi}(\theta, \varphi)$ are then used for the computation of the channel impulse response, e.g., in the equation (7.5-22) from [4].

A similar approach as described above can be also used when each antenna location corresponds to two antenna field patterns, e.g., for dual-polarized antennas.

In a channel model, multipath components can arrive at the UE with various polarizations. If the UE antenna were purely vertically polarized, for instance, any horizontally polarized signal component would be ignored (or attenuated heavily) by that antenna. Real smartphone antennas, however, will usually pick up energy in both polarizations to some extent (since they are not ideal polarized antennas).

The resulting polarization patterns, like in Fig. 5-a, show behavior that matches measured data from real phones. In practice, the polarization of the smartphone antennas is quite complex – not simply vertical or horizontal, especially when the phone is held. The model's rotated patterns successfully reproduce this complexity.

# Spatial Non-Stationarity

One of the most challenging aspects of modeling UE antennas is accounting for the user's effect – in particular, how the presence of a hand (or head or body) near the device influences the antenna performance. In channel modeling terms, this often falls under "blockage" or "shadowing" effects caused by the user.

The channel is not stationary over the UE antennas: some elements might be in a deep fade (blocked by the hand), while others are not. The Rel.19 model introduces a new near-field blockage model to capture this element-specific blocking behavior, replacing the older simplistic blockage approach.

Previously, 3GPP's channel model [2] included a so-called "Model A" for self-blocking. This model defined a single blockage region in terms of an angular sector relative to the device orientation. For example, in portrait mode, one might define that the user's head or hand blocks a cone of directions behind the phone as illustrated in Fig. 6-a.

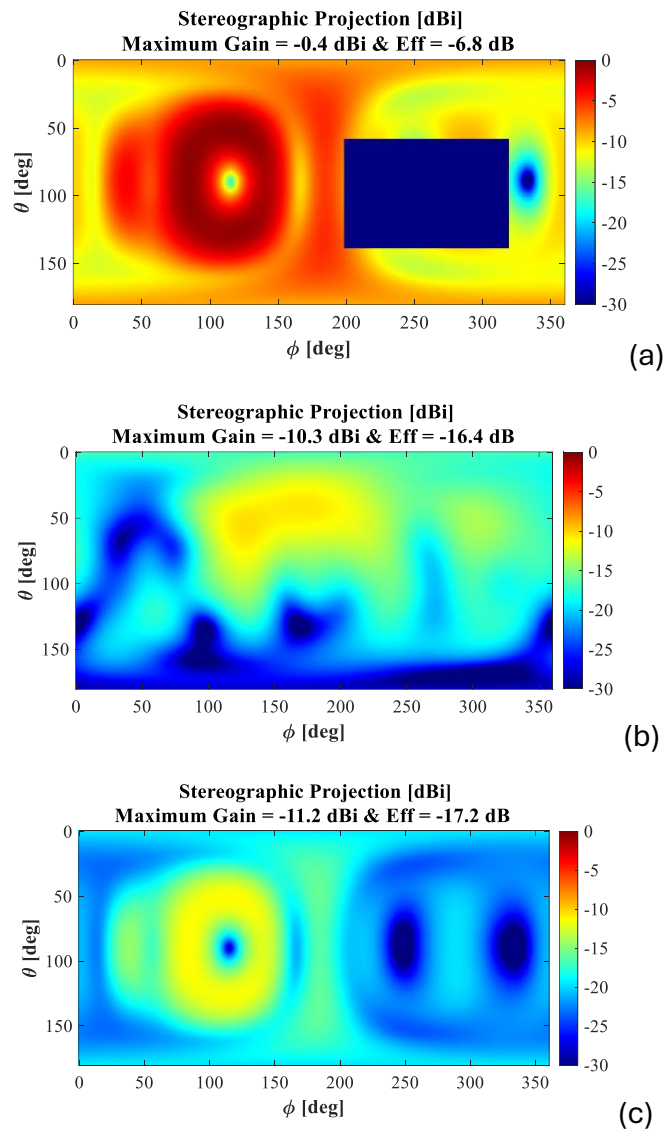

**FIGURE 6.** UE antenna blockage with 3GPP Model A (a), reference phone RF simulations (b), and new UE SNS model (c).

The Rel.19 update abandons the notion of a one-size-fits-all blockage cone. Instead, it takes an element-wise approach: each of the defined antenna positions on the UE is given its own attenuation factor to apply when the UE is in a certain blockage scenario. In the simulations, each UE is assigned a scenario with a certain probability defined in Tab. 7.6.14.2-1 in [4]. The probabilities were agreed as a consensus considering phone usage statistics reported in [11], Section 2.1.

For the frequencies up to 8.4 GHz, the attenuation values are derived from the simulation data of the reference smartphone with phantoms reported in [12]. The simulations were carried out in all blockage scenarios at different frequencies: 0.7, 2.0, 2.6, 3.8, 4.6 and 7.8 GHz. The antenna implementation and physical size requirements at frequencies below 1 GHz is typically different than for the higher frequencies. Therefore, below 1GHz the model is only applicable for 2-antenna system with antenna locations #4 and #8. In the range 1 - 8.4 GHz, the antenna attenuation was derived by averaging over frequency the values obtained at individual simulation points. An example of 10.8 dB attenuated free-space radiation pattern is shown in Fig. 6-c that is much closer to the real phantom simulation from Fig. 6-b than Model A shown in Fig. 6-a. Finally, the attenuation values for the frequencies around 15 GHz were based on measurements of the mockup phone with realistic hands and head fantoms report in [13], [14]. The resulting models are captured in Tab. 7.6.14.2-2 in [4].

In effect, the standard now takes in a default set of imbalance values so that anyone running simulations will automatically account for the fact that the UE's antennas are not receiving equal power when a user is holding the device. This is a step-change from earlier models. Note that the older stochastic Model A can still be used or extended for modeling body blockage (for example, the user's torso or another person causing shadowing of the signal path at a larger scale).

# Conclusion

The enhancements in the 3GPP Rel.19 UE antenna and blockage models significantly increase realism in simulations for handheld devices, enabling more accurate evaluation of future 6G technologies. By adopting realistic antenna patterns, orientations, and user blockage scenarios, these updated models help standardize system-level assessments, facilitating meaningful comparisons across the industry. Ultimately, these improvements support robust design decisions in both standardization and future 6G product development, bridging the gap between theoretical performance and real-world user experience.

# Biographies

**Simon Svendsen** (simon.svendsen@nokia.com) received his M.Sc.E.E. degree from Aalborg University, in 1995.

He has been working in the mobile industry since then. His career started at Bang & Olufsen Telecom, where his first task was to design antennas for DECT phones and verify the RF performance of the Front End. He has since worked for Maxon, Siemens, Motorola, and Molex doing mostly cellular antenna design for mobile phones including MCAD design, EM simulations, rapid prototyping and measurement. He joined Intel in January 2013, and was primarily working with cellular antenna designs for internal reference platforms and for customers. However, his focus in the last years at Intel shifted towards mmWave antenna design, and the RF FE architecture. He has been employed at Nokia since November 2018 and is working with standardization research for 3GPP 5G and 6G with focus on handheld devices. He has 50+ patents within the scope of antenna designs and 100+ patents within 3PGG standardization.

**Dimitri Gold** [Senior Member, IEEE] (dimitri.gold@nokia.com) received the M.Sc. degree (diploma with honors) in mathematical physics from M.V. Lomonosov Moscow State University, Russia, in 2007, and pursued Ph.D. studies in mathematical modeling and computational methods at the same university from 2007 to 2010. He received the Ph.D. degree (excellent grade) in mathematical information technology from the University of Jyväskylä, Finland, in 2012, and the title of Docent (Adjunct Professor) in wireless networking technology from the same university in 2016. His research interests include AI-native wireless networks, signal processing and propagation, and advanced simulation methods.

He is currently a 3GPP RAN1 delegate at Nokia Standards, Espoo, Finland, focusing on 6G research topics, particularly AI/ML for the physical layer and mobility. From 2020 to 2024, he served as a 3GPP RAN4 Delegate and lead for such topics as high-speed train (HST), integrated access and backhaul (IAB), and AI/ML in RAN, contributing from Rel-16 to Rel-19 3GPP studies. From 2019 to 2020, he was a senior researcher at Nokia Bell Labs, where he worked on AI-based network automation. He has also held positions as project manager and analyst at several companies and as a university lecturer and researcher. He has coauthored more than 40 academic publications and has submitted over 100 standard-related inventions.

Dr. Gold has served as Chair (2020–2022) and Vice-Chair (2018–2020) of the IEEE Finland Section and as a Chair of the joint chapter of Signal Processing/Circuit and Systems (2016-2018). He has received several recognition awards from IEEE and Nokia.

**Christian Rom** (christian.rom@nokia.com) M.Sc.EE. degree, in 2003 specialized in digital communications and CDMA receivers, and the Ph.D. degree in wireless communications in 2008 from Aalborg University. His Ph.D. research focused on physical layer parameter and algorithm study in a downlink OFDM-LTE context. Before joining Nokia, Christian worked as the RF Innovation Manager at Intel Denmark with focus on mmWave beamforming in smartphones and production. He has 20 years' experience in the wireless industry (Nokia/Intel/Infineon) with multiple roles in the modem chip industry for tier1 customer. During his career, he led large funding research projects in areas such as real world modelling and virtualization of propagation via ray-tracing to comparison of the models results with on-device measurements in the field, then linking these to link and system level tool development, PHY & MIMO Tx-Rx algorithms, and performance benchmarking and differentiation all the way to final mass scale product ramp up. He joined Nokia-Bell Laboratories Aalborg, Denmark, in January 2019. In 2022, he was awarded with the Outstanding Inventor in SIP cellular standards at Nokia.

**Volker Pauli** (volker.pauli@nokia.com) received the Dipl.Ing. and Dr.Ing. degrees (Hons.) from the University of Erlangen–Nuremberg, Germany, in 2003 and 2007, respectively. During this time, he was a Visiting Researcher at the École Polytechnique Fédérale de Lausanne (EPFL), Switzerland, and The University of British Columbia (UBC), Vancouver, Canada.

From 2007 until 2020, he was with Nomor Research GmbH, where for many years he was a Leader of the Research and Development group for system-level simulations of various wireless communication systems (3G, 4G, 5G, DVB-NGH, and WiFi), since 2008. Since 2020 he is with Nokia working filling various roles from in mobile communications research and simulator development. He has published more than 30 conference papers, journal articles, and white papers, and filed several patent applications. Over the years, he has been in touch with countless topics that appeared in 3GPP standardization.

Dr. Pauli's research interests are focused on layers one and two of mobile communication systems and link-to-system-level modelling for computationally efficient implementation of system-level simulators for mobile communication networks. He also works as an independent consultant, a trainer, and a technical expert for several organizations.

**Vuokko Nurmela, née Vuorinen** (vuokko.nurmela@nokia.com) received her M.Sc. degree in physics from the University of Helsinki in 1998.

She has been working for Nokia since 1997, contributing to mobile communication generations 2G, 3G, 4G, 5G and 6G. Her main areas of expertise are multiantenna algorithms and radio propagation in the wireless channel. She has been leading and conducting several measurement campaigns to investigate how radio waves behave between the transmitter and receiver. Currently she holds the position of Radio Propagation Owner in Nokia Mobile Networks in Espoo, Finland, where her responsibility is to promote better propagation understanding to support product decisions.

Mrs. Nurmela has 10 granted patents, and she has authored tens of conference and journal papers and 3GPP contributions. She has been awarded the title of Distinguished Member of Technical Staff in 2025.